\documentclass[12pt]{article}
\usepackage{amsfonts}
\usepackage[margin=2cm,nohead]{geometry}

\usepackage[normalem]{ulem}

\newcommand{\comment}[1]{}

\newcommand{\braket}[2]{{\langle {#1}\!\mid\!{#2} \rangle}}
\newcommand{\Hilbert}{{\cal H}}

\newtheorem{theorem}{Theorem}[section]
\newtheorem{definition}{Definition}[section]
\newtheorem{lemma}{Lemma}[section]

\newtheorem{property}{Property}[section]
\newtheorem{example}{Example}[section]
\newtheorem{remark}{Remark}[section]

\newcommand{\mat}{\left( \!\! \begin{array}{cc}}
\newcommand{\rix}{\end{array} \!\! \right)}

\newcommand{\Endproof}{\hfill$\Box$\\}

\usepackage{color}
\def\bR{\begin{color}{red}}
\def\bB{\begin{color}{blue}}
\def\bM{\begin{color}{magenta}}
\def\bC{\begin{color}{cyan}}
\def\bW{\begin{color}{white}}
\def\bBl{\begin{color}{black}}
\def\bG{\begin{color}{green}}
\def\bY{\begin{color}{yellow}}
\def\ec{\end{color}\ }

\usepackage{xy}
\xyoption{matrix}
\xyoption{frame}
\xyoption{arrow}
\xyoption{arc}

\usepackage{ifpdf}
\ifpdf
\else
\PackageWarningNoLine{Qcircuit}{Qcircuit is loading in Postscript mode.  The Xy-pic options ps and dvips will be loaded.  If you wish to use other Postscript drivers for Xy-pic, you must modify the code in Qcircuit.tex}
\xyoption{ps}
\xyoption{dvips}
\fi

\entrymodifiers={!C\entrybox}

\newcommand{\ket}[1]{{\left\vert{#1}\right\rangle}}
\newcommand{\qw}[1][-1]{\ar @{-} [0,#1]}
\newcommand{\qwx}[1][-1]{\ar @{-} [#1,0]}
\newcommand{\cw}[1][-1]{\ar @{=} [0,#1]}
\newcommand{\cwx}[1][-1]{\ar @{=} [#1,0]}
\newcommand{\gate}[1]{*+<.6em>{#1} \POS ="i","i"+UR;"i"+UL **\dir{-};"i"+DL **\dir{-};"i"+DR **\dir{-};"i"+UR **\dir{-},"i" \qw}
\newcommand{\meter}{*=<1.8em,1.4em>{\xy ="j","j"-<.778em,.322em>;{"j"+<.778em,-.322em> \ellipse ur,_{}},"j"-<0em,.4em>;p+<.5em,.9em> **\dir{-},"j"+<2.2em,2.2em>*{},"j"-<2.2em,2.2em>*{} \endxy} \POS ="i","i"+UR;"i"+UL **\dir{-};"i"+DL **\dir{-};"i"+DR **\dir{-};"i"+UR **\dir{-},"i" \qw}

\newcommand{\control}{*!<0em,.025em>-=-<.2em>{\bullet}}
\newcommand{\controlo}{*+<.01em>{\xy -<.095em>*\xycircle<.19em>{} \endxy}}
\newcommand{\ctrl}[1]{\control \qwx[#1] \qw}
\newcommand{\ctrlo}[1]{\controlo \qwx[#1] \qw}

\newcommand{\qswap}{*=<0em>{\times} \qw}

\newcommand{\gategroup}[6]{\POS"#1,#2"."#3,#2"."#1,#4"."#3,#4"!C*+<#5>\frm{#6}}

\newcommand{\lstick}[1]{*!R!<.5em,0em>=<0em>{#1}}
\newcommand{\ustick}[1]{*!D!<0em,-.5em>=<0em>{#1}}

\newcommand{\Qcircuit}{\xymatrix @*=<0em>}

\begin{document}

\title{Quantum Hashing}

\author{Farid Ablayev\thanks{Kazan Federal University}  \and  Alexander Vasiliev\thanks{Kazan Federal University}}

\date{}


\maketitle

\begin{abstract}

We present  a version of quantum hash function based  on  non-binary discrete
functions. The proposed quantum procedure  is ``classical-quan\-tum'', that is,
it  takes a  classical bit string  as an input and produces a quantum state.
The resulting function   has the property of a one-way function  (pre-image
resistance), in addition it has the properties analogous to classical
cryptographic hash second pre-image resistance and collision resistance.

This function can be naturally  used in a quantum digital signature
protocol.

\end{abstract}



\section{Preliminaries and introduction}

Hashing has a lot of fruitful applications in  computer science, in particular
the public-key cryptography relies on cryptographic hash functions.
Hash functions are designed to take a string of large  length (theoretically
any length) as an input and produce a short (in practice a fixed-length) hash
value. A cryptographic hash function must be additionally able to withstand all
known types of cryptanalytic attack. At least, it must have the following
properties:
\begin{itemize}
\item {{\em Pre-image resistance}} (or equivalently First pre-image resistance).

Given a hash $v$ it should be ``computationally  difficult to invert'' hash
function $hash$, that is, to find any message $w$ such that $v = hash(w)$.
The  pre-image resistance property together  with the ``easy computation''
property (given $w$ it is easy to compute a value  $v = hash(w)$)   is
known as the {\em one-way property}.

\item {\em Second pre-image resistance.}

Given an input $w$ it should be ``computationally difficult'' to find
another input $w'$ such that  $w\not= w'$ and $hash(w) = hash(w')$.
Functions that lack this property are vulnerable to second pre-image
attacks.

\item   {\em  Collision resistance.}

It should be ``computationally difficult'' to find two different messages $w$ and $w'$ such that $hash(w) = hash(w')$.
Such a pair is called a cryptographic hash collision.
This property is sometimes referred to as strong collision resistance.
\end{itemize}
The ``computationally difficult (hard) problem'' means that for the problem
considered there must be no algorithm (oriented for realization in realistic
computational model) except  enumeration algorithm of possible instances that
potentially fit the problem solution. Classical cryptographic functions rely on
hardness of certain mathematical problems, such as integer factorization and
discrete logarithm. Besides these well known problems several  other
potentially hard problems were discovered  and such investigations are still in
progress.

The main problem arrives in this aspect is to proof for a  certain candidate,  that considered  problem is  really hard.
However, proving for a particular function the  one-way property would imply that $P \not= NP$. The latest problem is a modern mathematical challenge of era.


In contrast  to classical approach quantum cryptography is
based on foundations of quantum mechanics and information properties of quantum
systems. In the fall of the last century and  last decades several models  of
quantum one-way functions were proposed. In
\cite{Kashefi-Kerenidis:2007:Quantum-One-Way} a family of
``classical-classical'' functions was considered, whose inputs and outputs are
classical binary strings. These functions are candidates to be hardly
invertible not only classically but also quantumly.  Authors call such
functions  quantum one-way functions.

Quantum one-way functions defined by Gottesman and Chuang \cite{GC:2001:Quantum-Digital-Signatures} are ``classical-quantum'' one-way functions, that is, such a function  takes a  classical bit string as an input and produces a quantum state. Another type of ``classical-quantum'' one-way function was invented by Buhrman et al. \cite{Buhrman:2001:Fingerprinting} 
based on binary error-correcting code and is known as quantum fingerprinting.
Based on classical-quantum notion of a quantum one-way function
several schemes of quantum digital signature were proposed
\cite{GC:2001:Quantum-Digital-Signatures,Lu-Feng:2005:signature-quantum-one-way-based,Zhou-et-al:2011:proxy-signature}.

In this research paper we define a notion of ``classical-quantum''
hashing function which is a natural extension of the notion of ``classical-quantum''
one-way function. We present a non-binary variant of quantum hashing function and prove
its cryptographic properties. We propose an effective  computation scheme
of the quantum hashing function based on the quantum branching program model.
Finally, as an application of quantum hashing functions we present a digital signature scheme.

{\em Organization of the paper.}  We start the paper with the discussion of
quantum one-way function definition and we come to the notion of quantum hash
function (section \ref{quantum-hash-functions}). In the section \ref{Quantum-Fingerprinting} we show that  known
fingerprinting function \cite{Buhrman:2001:Fingerprinting} is a quantum hashing
function. In the section \ref{Quantum-Hashing} we present our variant of quantum hashing
function and prove that it has a desired hashing  property.

The section is devoted to the application of quantum hash functions
for a digital signature scheme.

\section{Quantum one-way and hashing functions}\label{quantum-hash-functions}

The definition of a quantum one-way function is based on
\cite{GC:2001:Quantum-Digital-Signatures} and explicitly presented in
\cite{Lu-Feng:2005:signature-quantum-one-way-based,Zhou-et-al:2011:proxy-signature}.
Let
\begin{equation}
 \psi : \{0,1\}^n \to  (\Hilbert^2)^{\otimes s}
 \end{equation}
be a function (classical-quantum function), where
\[(\Hilbert^2)^{\otimes s}=\Hilbert^2\otimes\dots \otimes\Hilbert^2 = \Hilbert^{2^s} \]
is a $2^s$-dimensional  Hilbert space   made up of $s$ copies of a single qubit space $\Hilbert^2$.

We will also  use notation
 \begin{equation}
  \psi : w \mapsto  \ket{\psi(w)}
 \end{equation}
 for $\psi$, which is frequently used in different papers.

\begin{itemize}
 \item Function $\psi$ is called a quantum one-way if it is
\begin{itemize}
 \item easy to compute: there is quantum polynomial-time algorithm that on input $w$ outputs $\ket{\psi(w)}$.
 \item  hard to invert: given $\ket{\psi(w)}$, it is impossible to invert $w$
by virtue of fundamental quantum information theory.
\end{itemize}
\end{itemize}

\begin{property}
If $n\gg s$ in the definition above,  then  given $\ket{\psi(w)}$, it is impossible to obtain $w$.
 \end{property}\label{one-way}
{\em Proof.}  This pre-image resistance property follows from Holevo bound
\cite{Holevo:1973:bound}, since no more than $O(s)$ classical bits of information can be
extracted from $s$ qubits and the original message contains  $n\gg s$ bits.
\Endproof

\begin{example}[One-way function]
 A word $w\in \{0,1\}^n$ is encoded by a single qubit:
\[ \psi : w \mapsto  \cos\left(\frac{2\pi w}{2^n}\right)\ket{0} +
  \sin\left(\frac{2\pi w}{2^n}\right)\ket{1}.  \]
Here we   treat $w=w_0\dots w_{n-1}$ also as a number
    $w=w_0+w_12^1+\dots +w_{n-1}2^{n-1}$.
\end{example}

Clearly, we have   that $\psi$  has the one-way property of the definition and
the Property \ref{one-way} above. What we need in extra and what is implicitly
assumed  in various papers (see for example \cite{Lu-Feng:2005:signature-quantum-one-way-based,Zhou-et-al:2011:proxy-signature}) is
a collision resistance property. However, there is still no such notion as
\emph{quantum collision}. The reason why we need to define it is the observation
that in quantum hashing there might be no collisions in the classical sense:
since quantum hashes are quantum states they can store arbitrary amount of data and
can be different for unequal messages. But the procedure of comparing those quantum
states implies measurement, which can lead to collision-type errors.

So, a \emph{quantum collision} is a situation when a procedure that tests an
equality of quantum hashes outputs true, while hashes are different. This
procedure can be a well-known SWAP-test \cite{Buhrman:2001:Fingerprinting} or
something that is adapted for specific hashing function. Anyway, it deals
with the notion of distinguishability  of quantum states. And since
non-orthogonal quantum states cannot be perfectly distinguished, we require
them to be ``nearly orthogonal''.

To formalize the notion of ``nearly orthogonality'' we will call states
$\ket{\psi_1}$ and $\ket{\psi_2}$ \emph{$\delta$-orthogonal} if
$$
\left|\braket{\psi_1}{\psi_2}\right| < \delta.
$$

Thus, for a quantum hash-function it is important to have an ability to reliably
compare quantum hashes of different words  and those
quantum  states need to be distinguishable with high probability, that is, they
have to pass non-equality tests.

\paragraph{REVERSE-test.}

Whenever we need to check if a quantum state $\ket{\psi(w)}$ is a hash of a
classical message $v$,
one can use the procedure that we call a \emph{Reverse-test} (the idea of such test
for the case of a quantum message given by $\ket{v}$ was described in
\cite{GC:2001:Quantum-Digital-Signatures}, but it had not been given its own name).

Essentially the test applies the procedure that inverts the creation of a quantum
hash, i.e. it ``uncomputes'' the hash to the initial state (usually the all-zero state).

Formally, let the procedure of quantum hashing of message $w$ consist of unitary
transformation $U(w)$, applied to initial state $\ket{0}$, i.e. $\ket{\psi(w)}=U(w)\ket{0}$.
Then the Reverse-test, given $v$ and $\ket{\psi(w)}$, applies $U^{-1}(v)$ to
the state $\ket{\psi(w)}$ and measures the resulting state. It outputs $v=w$ iff the measurement
outcome is $\ket{0}$. So, if $v=w$, then
$U^{-1}(v)\ket{\psi(w)}$ would always give $\ket{0}$, and Reverse-test would give
the correct answer. Otherwise, the resulting state would be $\delta$-orthogonal to
$\ket{0}$ since unitary operators preserve inner product:
$$
(\ket{0}, U^{-1}(v)\ket{\psi(w)}) = (U^{-1}(v)\ket{\psi(v)}, U^{-1}(v)\ket{\psi(w)}) =
(\ket{\psi(v)}, \ket{\psi(w)}) = \braket{\psi(v)}{\psi(w)} < \delta.
$$

Overall, for this test has one-sided error bounded by $\delta$ if quantum hashes of
different messages are $\delta$-orthogonal.

\paragraph{SWAP-test.}

A more general test, that checks the equality of two arbitrary states is a well-known
SWAP-test \cite{Buhrman:2001:Fingerprinting}, given by the following circuit:
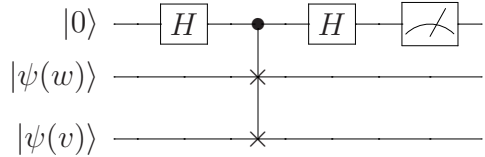
\[\quad\quad\quad
\Qcircuit  @C=0.75em @R=1.0em {
\lstick{\ket{0}} & \qw & \gate{H} & \qw & \ctrl{1} & \qw & \gate{H} & \qw & \meter & \qw\\
\lstick{\ket{\psi(w)}} & \qw & \qw & \qw & \qswap \qwx[2] & \qw & \qw & \qw & \qw & \qw\\
\\
\lstick{\ket{\psi(v)}} & \qw & \qw & \qw & \qswap & \qw & \qw & \qw & \qw & \qw\\
}
\]

Applied to quantum hash codes it outputs $\ket{\psi(w)}=\ket{\psi(v)}$, if the measurement
result of the first qubit is $\ket{0}$.

\begin{property}\label{SWAP-test-error-probability} The probability of obtaining $\ket{0}$ in the SWAP-test is equal to $\frac{1}{2}\left(1+|\braket{\psi(w)}{\psi(v)}|^2\right)$.
\end{property}

{\em Proof.} See \cite{Buhrman:2001:Fingerprinting}.
\Endproof

The probability of error of the SWAP-test inherently depends on the value of
the inner product of $\ket{\psi(w)}$ and $\ket{\psi(v)}$ -- it is minimal
(close to 1/2), when these states are ``nearly orthogonal'' \cite{GC:2001:Quantum-Digital-Signatures}.

Thus, the property of being $\delta$-orthogonal for quantum states is
crucial for quantum collision resistance
\cite{GC:2001:Quantum-Digital-Signatures}. And at this point we come to a
notion of a $\delta$-resistance.

\begin{definition}[$\delta$-resistance]\label{delta-collision-resistance}
We call a function $\psi : w\mapsto\ket{\psi(w)}$ $\delta$-resistant if
for any pair of inputs $w$, $w'$, $w\neq w'$ their images are
$\delta$-orthogonal:
$$
\left|\braket{\psi(w)}{\psi(w')}\right| < \delta.
$$
\end{definition}

$\delta$-resistance of a hash function is a key property for bounding the probability of error for SWAP-test and
REVERSE-test.
Note, that the ideas of $\delta$-resistance and REVERSE-test were used in \cite{Buhrman:2001:Fingerprinting}.

Note, that this $\delta$-resistance property also corresponds to the classical \emph{Second pre-image
resistance}, since we cannot find two different messages for which the SWAP-test would erroneously
output true with probability close to 1.

Finally, we naturally  come to the following  definition of classical-quantum hash function.
\begin{definition}[$(n,s,\delta)$-quantum hash function]\label{quantum-hash-function}
We call a function
 \[ \psi : \{0,1\}^n \to  (\Hilbert^2)^{\otimes s} \]
 $(n,s,\delta)$-quantum hash function, if it is quantum  one-way and $\delta$-resistant function.

\end{definition}

The following property is an immediate implication of Definition
\ref{quantum-hash-function} and Property \ref{SWAP-test-error-probability}.

\begin{property}
If a function $\psi : w\mapsto\ket{\psi(w)}$ is $(n,s,\delta)$-quantum hash function,
then the SWAP-test distinguishes the hashes of two messages $w\neq w'$
with probability $\frac{1}{2}(1-\delta^2)$.
\end{property}
{\em Proof.} By the Property \ref{SWAP-test-error-probability} the probability of error for
the SWAP-test is $\frac{1}{2}\left(1+|\braket{\psi(w)}{\psi(w')}|^2\right)$.
Since for $\delta$-resistant function $\left|\braket{\psi(w)}{\psi(w')}\right| < \delta$,
this test distinguishes quantum hashes of the pair of messages $w\neq w'$ with
probability $\frac{1}{2}(1-\delta^2)$.
\Endproof

\begin{remark}
The error probability of the SWAP-test can be reduced to any $\epsilon > 0$ by standard repetition technique,
that is by performing this test upon $k=O(\log 1/\epsilon)$ copies of compared states.
In other words, we could have used a function \[ \psi' : {0,1}^n \mapsto
\ket{\psi'(u)}, \] given by
\[\ket{\psi'(u)}=\ket{\psi(u)}^{\otimes k}= \ket{\psi(u)}\otimes \dots \otimes \ket{\psi(u)}\]
 In this case, the total number of qubits to encode a word of
length $n$ is
$ O(\log n\log(1/\epsilon))$.
\end{remark}

In the next two sections we show that  known quantum fingerprinting function is
a quantum hashing function and we present our construction of quantum hash
function with a slightly different characteristics.

\section{Quantum Fingerprinting}\label{Quantum-Fingerprinting}

In \cite{Buhrman:2001:Fingerprinting} Buhrman et al. defined a quantum one-way function
\[ f_E: u\mapsto \ket{f_E(u)}\]
 of a bit string $u\in\{0,1\}^n$, which they have called \emph{quantum fingerprinting}. Based on the existence of the binary error-correcting code
$E : \{0, 1\}^n \mapsto \{0,1\}^m $ with $m=cn$ and Hamming distance $(1-\delta)m$ (for $c>2$ and $\delta<9/10 + 1/(15c)$) they have defined a quantum
fingerprint of $u$ as follows:
\[ \ket{f_E (u)} = \frac{1}{\sqrt{m}}\sum_{i=1}^m(-1)^{E_i(u)}\ket{i}  \]
where $E_i(u)$ denotes the $i$-th bit of $E(u)$.


\begin{property} For a $\delta \approx  9/10 + 1/(15c)$ the quantum fingerprinting function  $f_E$ is an $(n, O(\log n), \delta)$-quantum hashing function.
\end{property}
{\em Proof.} The function $f_E$ is the quantum  one-way and is   $\delta$-orthogonal for the $\delta \approx 9/10 + 1/(15c)$ \cite{Buhrman:2001:Fingerprinting}.
\Endproof


\section{Quantum Hashing}\label{Quantum-Hashing}

In this section we propose a quantum hashing function based on construction from \cite{ablayev-vasiliev:2009:EPTCS}.

Let $N=2^n$. Let $K=\{k_i: k_i\in\{0,\ldots,N-1\}\}$ and $d=|K|$. We define a classical-quantum function
\[ h_K : \{0,1\}^n \to ({\mathbb C}^2)^{\otimes (d+1) } \]
as follows. For a message $M\in\{0,1\}^n$ we let
%
\[
\ket{h_K(M)} =
\frac{1}{\sqrt{d}}\sum\limits_{i=1}^d\ket{i}
\left(\cos\frac{2\pi k_i
M}{N}\ket{0}+\sin\frac{2\pi k_i M}{N}\ket{1}\right).
\]
\begin{theorem}\label{fingerprinting-is-hashing} For arbitrary $\epsilon >0$ there exists  a set $K$ with $|K|=\lceil(2/\epsilon^2)\ln(2N)\rceil$  such that quantum function $h_K$ is an $(n, O(\log n +\log 1/\epsilon), \epsilon)$-quantum hashing function.
\end{theorem}

\subsection{Proof of the Theorem \ref{fingerprinting-is-hashing}}

To prove the Theorem \ref{fingerprinting-is-hashing} we will show that the function
$h_K$ is a one-way function and has $\delta$-resistance.

\subsubsection{$\delta$-resistance of $h_K$}

To prove $\delta$-resistance $h_K$
we recall some definitions and statements that we will use later.

\begin{definition}
The discrete Fourier transform of the characteristic function of a set
$K\subset \mathbb{Z}_N$ is the function
$$f_K(l) = \sum\limits_{k\in K} e^{i\frac{2\pi k l}{N}}.$$
\end{definition}

Let $\lambda(K) = \max\limits_{l\neq0}\frac{|f_K(l)|}{|K|}$ and let
$\delta(K)=\max\limits_{l\neq0}\frac{|\mathbf{Re}(f_K(l))|}{|K|}$. As it was
mentioned in \cite{RSW:1993:Small-uniform-sets}, $\lambda(K)$ gives some
measure of randomness of the set $A$: the smaller it is, the more ``random'' A
is. $\delta(K)$ is defined similarly, but it uses the real part of $f_K(l)$.
Clearly, $\delta(K)\leq\lambda(K)$.

For our technique will need a set $K$ with $\delta(K)$ as small as possible. In
\cite{RSW:1993:Small-uniform-sets} a construction is given that for
$\epsilon=\left(\frac{1}{\log{N}}\right)^{O(1)}$ yields a set $K$ with
$|K|=(\log{N})^{O(1)}$ and $\lambda(K)\leq\epsilon$.

Additionally, in \cite{ablayev-vasiliev:2008:ECCC} a proof of the following
lemma is given.

\begin{lemma}\label{existence-of-a-good-set}
For any $\varepsilon\in(0,1)$ there exists a set $K$ with
$|K|=\lceil\frac{2}{\varepsilon^2}\ln(2N)\rceil$ and $\delta(K)<\varepsilon$.
\end{lemma}

Note, that in \cite{ablayev-vasiliev:2008:ECCC} we did not use the notation
$\delta(K)$, the sum $\frac{1}{|K|}\sum\limits_{k\in K} cos\frac{2\pi k l}{N}$
was used instead.

Let $\delta\in(0,1)$ and pick a set $K$ satisfying Lemma \ref{existence-of-a-good-set} for $\varepsilon = \delta^2$. In this case
$h_K$ is $\delta$-resistant, since for any pair of messages $M_1\neq M_2$
$$
\left|\braket{h_{M_1}\,}{h_{M_2}}\right| =
\left|\frac{1}{|K|}\sum\limits_{i=1}^{|K|} \left(\cos\frac{2\pi
k_i M_1}{N}\cos\frac{2\pi k_i M_2}{N} +\sin\frac{2\pi
k_i
M_1}{N}\sin\frac{2\pi k_i M_2}{N}\right)\right| = $$
$$
\left|\frac{1}{|K|}\sum\limits_{i=1}^{|K|}
\cos\frac{2\pi k_i (M_1-M_2)}{N}\right| \leq \delta(K)<\delta.
$$

Thus, $h_K$ hashes $n$-bit messages into quantum states of
$O(\log{n}+\log 1/{\delta^2})$ qubits and provides $\delta$-orthogonality of
quantum hashes.



\subsubsection{$h_k$ is a quantum one-way function}

\paragraph{Irreversibility.}

Since to hash an $n$-bit message we use about $O(\log n)$ qubits, then by Holevo bound no more than
$O(\log n)$ bits of information can be extracted from it, i.e. one cannot restore all of $n$ bits.

\paragraph{Effective computation.}

The proposed hashing function can be efficient implemented in
quantum computational models that allow classical control, such as Quantum
Branching Programs \cite{Ablayev-Gainutdinova-Karpinski:2001:QBP}.

Below is a read-once quantum branching program (a quantum OBDD) that hashes an $n$-bit
string $M=b_1b_2\ldots b_n$ into $\ket{h_K(M)}$ using $O(\log{n})$ qubits:

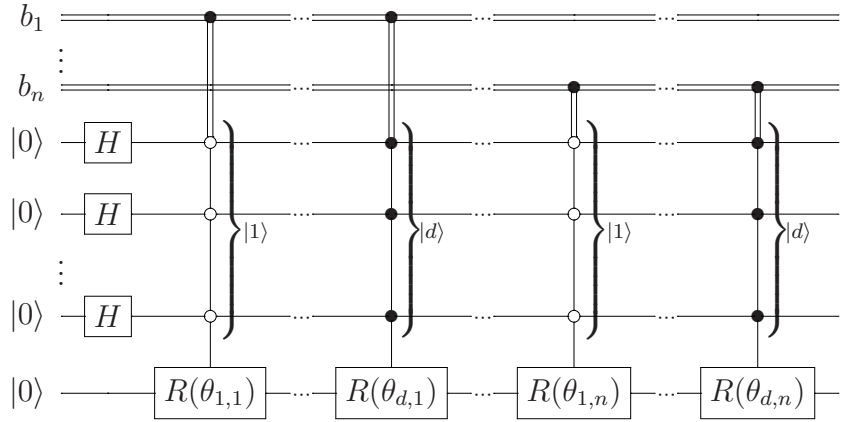
\[\quad\quad\quad
\Qcircuit @C=0.75em @R=1.0em {
\lstick{b_1} & \cw & \control\cw\cwx[3] & \cw & ~_{\cdots}\quad & \control\cw\cwx[3] & \cw & ~_{\cdots}\quad & \cw & \cw & ~_{\cdots}\quad & \cw & \cw \\
\vdots \\
\lstick{b_n} & \cw & \cw & \cw & ~_{\cdots}\quad & \cw & \cw & ~_{\cdots}\quad & \control\cw\cwx[1] & \cw & ~_{\cdots}\quad & \control\cw\cwx[1] & \cw \\
\lstick{\ket{0}} & \gate{H} & \ctrlo{1} & \qw  & ~_{\cdots}\quad & \ctrl{1} & \qw & ~_{\cdots}\quad & \ctrlo{1} & \qw  & ~_{\cdots}\quad & \ctrl{1} & \qw\\
\lstick{\ket{0}} & \gate{H} & \ctrlo{2} & \qw  & ~_{\cdots}\quad & \ctrl{2} &  \qw & ~_{\cdots}\quad & \ctrlo{2} & \qw  & ~_{\cdots}\quad & \ctrl{2} & \qw\\
\vdots &&\ustick{\quad\quad\quad^{\ket{1}}} &&& \ustick{\quad\quad~~^{\ket{d}}}&&& \ustick{\quad\quad\quad^{\ket{1}}}&&& \ustick{\quad\quad~~^{\ket{d}}}\gategroup{4}{3}{7}{3}{1em}{\}}\gategroup{4}{6}{7}{6}{1em}{\}}\gategroup{4}{9}{7}{9}{1em}{\}}\gategroup{4}{12}{7}{12}{1em}{\}} \\
\lstick{\ket{0}} & \gate{H} & \ctrlo{1} & \qw & ~_{\cdots}\quad & \ctrl{1}& \qw & ~_{\cdots}\quad & \ctrlo{1} & \qw  & ~_{\cdots}\quad & \ctrl{1} & \qw\\
\lstick{\ket{0}}& \qw & \gate{R(\theta_{1,1})} & \qw & ~_{\cdots}\quad & \gate{R(\theta_{d,1})} & \qw & ~_{\cdots}\quad & \gate{R(\theta_{1,n})} & \qw & ~_{\cdots}\quad & \gate{R(\theta_{d,n})} & \qw\\
}
\]

Here, $R(\theta_{i,j})$ denotes a rotation by the angle $\frac{4\pi k_i
2^j}{N}$ around the $\hat{y}$ axis of the Bloch sphere, and $d=|K|$.

%


\subsection{Numerical Results}

The aforementioned methods for computing the set $K$ of hashing parameters
make sense for comparatively large $N$. For smaller $N$ the influence of
$\epsilon$ results in a quite big sizes of the set $K$. This problem is
especially important for quantum digital signature protocol, where the value
of $N$ is not very large.

To deal with this problem we have developed a genetic algorithm that gives
good results in acceptable time. Here are some examples of its work:

{\tiny
\begin{tabular}{|c|c||c|l|}
  \hline
  $N$ & $d=|K|$ & $\epsilon<0.01$ & $K$ \\
  \hline
32 & 15 & 0,0039 & \{1, 2, 3, 4, 5, 6, 7, 9, 10, 11, 12, 13, 14, 15, 16\}\\
64 & 33 & 0,0067 & \{3, 8, 10, 11, 11, 12, 16, 20, 21, 22, 23, 24, 28, 29, 30, 31, 31, \\&&&
                    35, 36, 37, 42, 45, 46, 45, 47, 50, 52, 56, 57, 60, 61, 62, 63\}\\
128 & 33 & 0,0099 & \{2, 6, 14, 18, 19, 23, 28, 29, 31, 32, 33, 34, 40, 47, 54, 56, 60, \\&&&66, 70, 75, 76, 77, 77, 80, 84, 87, 87, 92, 102, 107, 111, 115, 116\}\\
256 & 65 & 0,0087 & \{1, 2, 2, 8, 14, 23, 23, 23, 31, 35, 40, 42, 45, 46, 49, 55, 58,\\&&&
59, 60, 63, 66, 73, 73, 76, 77, 78, 83, 83, 88, 89, 99, 100, 101,\\&&&
104, 106, 108, 127, 134, 140, 141, 145, 151, 160, 160, 162, 172,\\&&&
174, 179, 182, 186, 187, 190, 204, 215, 216, 217, 219, 220, 233,\\&&&
234, 235, 238, 240, 241, 245\}\\
512 & 65 & 0,0094 & \{1, 28, 34, 38, 38, 39, 46, 49, 50, 58, 59, 69, 81, 97, 99, 103,\\&&&
108, 118, 123, 132, 156, 168, 177, 181, 197, 198, 200, 203, 204,\\&&&
220, 229, 230, 233, 251, 269, 284, 301, 301, 303, 304, 335, 337,\\&&&
349, 351, 352, 353, 365, 368, 381, 393, 397, 398, 400, 412, 417,\\&&&
436, 437, 451, 459, 467, 476, 478, 483, 488, 509\}\\
1024 & 65 & 0,0100 & \{13, 42, 50, 60, 63, 63, 65, 116, 122, 127, 141, 159, 162, 184,\\&&&
185, 189, 190, 201, 223, 263, 270, 272, 276, 306, 324, 334, 339,\\&&&
345, 346, 353, 354, 363, 379, 394, 410, 414, 433, 445, 447, 472,\\&&&
493, 494, 505, 507, 565, 567, 572, 590, 609, 707, 739, 770, 781,\\&&&
809, 812, 814, 819, 851, 855, 878, 893, 900, 958, 960, 998\}\\
2048 & 129 & 0,0069 & \{5, 26, 40, 54, 65, 80, 81, 102, 113, 115, 148, 168, 178, 190,\\&&&
216, 221, 236, 263, 263, 288, 300, 313, 314, 314, 380, 388, 423,\\&&&
426, 428, 439, 443, 459, 461, 492, 494, 529, 546, 547, 575, 596,\\&&&
608, 619, 620, 652, 653, 684, 685, 686, 739, 744, 746, 748, 758,\\&&&
773, 787, 813, 835, 846, 849, 854, 860, 865, 906, 927, 946, 992,\\&&&
1014, 1061, 1076, 1086, 1091, 1105, 1116, 1122, 1134, 1141, 1143,\\&&&
1160, 1168, 1208, 1215, 1252, 1262, 1271, 1287, 1310, 1328, 1369,\\&&&
1373, 1393, 1423, 1432, 1439, 1454, 1459, 1484, 1504, 1505, 1535,\\&&&
1570, 1573, 1578, 1597, 1612, 1624, 1643, 1653, 1671, 1726, 1732,\\&&&
1745, 1764, 1766, 1770, 1775, 1797, 1818, 1850, 1850, 1856, 1862,\\&&&
1863, 1924, 1931, 1943, 1994, 2026, 2028, 2045\}\\
4096 & 129 & 0,0092 & \{153, 179, 186, 341, 366, 384, 445, 465, 469, 495, 533, 574, 587,\\&&&
593, 613, 678, 760, 771, 823, 851, 869, 878, 924, 995, 1057, 1072,\\&&&
1074, 1140, 1235, 1267, 1286, 1302, 1351, 1371, 1415, 1431, 1476,\\&&&
1494, 1539, 1598, 1677, 1685, 1710, 1718, 1765, 1796, 1845, 1871,\\&&&
1916, 1924, 1939, 1953, 1977, 1985, 2040, 2062, 2066, 2070, 2079,\\&&&
2093, 2224, 2261, 2338, 2385, 2415, 2508, 2534, 2538, 2564, 2583,\\&&&
2614, 2626, 2627, 2639, 2691, 2699, 2720, 2723, 2735, 2742, 2747,\\&&&
2791, 2821, 2823, 2847, 2952, 2973, 3002, 3005, 3066, 3070, 3071,\\&&&
3127, 3128, 3138, 3186, 3189, 3293, 3383, 3393, 3408, 3414, 3415,\\&&&
3428, 3487, 3488, 3504, 3524, 3634, 3676, 3695, 3705, 3745, 3748,\\&&&
3752, 3772, 3810, 3849, 3851, 3878, 3888, 3913, 3925, 3949, 3995,\\&&&
4016, 4073, 4079, 4089\}\\
8192 & 129 & 0,0093 & \{53, 144, 144, 196, 311, 320, 384, 400, 464, 481, 624, 711, 767,\\&&&
789, 792, 871, 940, 1119, 1147, 1171, 1222, 1224, 1240, 1377,\\&&&
1543, 1605, 1669, 1743, 1746, 1824, 1927, 1928, 2099, 2115, 2293,\\&&&
2321, 2430, 2477, 2481, 2494, 2497, 2566, 2571, 2656, 2922, 2949,\\&&&
2975, 3064, 3105, 3151, 3184, 3297, 3388, 3395, 3514, 3733, 3907,\\&&&
3915, 3926, 3927, 3976, 4200, 4207, 4210, 4278, 4327, 4402, 4521,\\&&&
4668, 4724, 4744, 4763, 4764, 4814, 4916, 5047, 5070, 5090, 5130,\\&&&
5429, 5437, 5493, 5529, 5578, 5627, 5733, 5800, 5862, 5870, 5954,\\&&&
5989, 5991, 6038, 6041, 6076, 6138, 6179, 6204, 6229, 6297, 6331,\\&&&
6389, 6392, 6447, 6531, 6632, 6681, 6700, 6835, 6854, 6859, 7202,\\&&&
7393, 7496, 7655, 7699, 7715, 7742, 7745, 7763, 7790, 7864, 7900,\\&&&
7922, 7979, 8003, 8044, 8188, 8190\}\\
16384 & 129 & 0,0099 & \{11, 41, 216, 229, 303, 307, 326, 366, 666, 672, 1083, 1097,\\&&&
1151, 1184, 1246, 1341, 1361, 1389, 1425, 1457, 1558, 1707, 1977,\\&&&
2038, 2188, 2223, 2281, 2287, 2608, 2933, 2961, 2994, 3065, 3091,\\&&&
3140, 3164, 3690, 3708, 3918, 3919, 4152, 4225, 4270, 4477, 4892,\\&&&
4903, 4919, 5264, 5307, 5337, 5558, 5723, 5829, 5910, 5984, 6020,\\&&&
6164, 6192, 6278, 6287, 6357, 6401, 6521, 6604, 6904, 6974, 7704,\\&&&
7766, 7830, 7947, 8015, 8049, 8280, 8710, 8901, 9034, 9098, 9212,\\&&&
9245, 9351, 9389, 9452, 9743, 10072, 10083, 10100, 10107, 10308,\\&&&
10497, 10552, 10576, 10585, 10655, 11032, 11176, 11284, 11315,\\&&&
11386, 11457, 11726, 11839, 11977, 11979, 12049, 12244, 12488,\\&&&
12844, 12860, 12897, 13092, 13313, 13349, 13443, 13483, 13988,\\&&&
14090, 14174, 14205, 14396, 14630, 15030, 15060, 15119, 15604,\\&&&
15987, 16002, 16029, 16164, 16316\}\\
32768 & 257 & 0,0068 & \{15, 136, 303, 598, 702, 817, 1147, 1150, 1225, 1849, 1934,\\&&&
2071, 2109, 2198, 2204, 2386, 2596, 2839, 2908, 2983, 3293,\\&&&
3423, 3434, 3859, 4078, 4326, 4438, 4632, 4641, 4811, 4865,\\&&&
4905, 4908, 5226, 5269, 5491, 5561, 5592, 5722, 6025, 6053,\\&&&
6119, 6233, 6322, 6351, 6446, 6508, 6544, 7177, 7368, 7480,\\&&&
7665, 7685, 7786, 7928, 8111, 8197, 8287, 8362, 8621, 8685,\\&&&
9077, 9104, 9140, 9161, 9400, 9547, 9548, 9600, 10030, 10303,\\&&&
10583, 10716, 10944, 10976, 11334, 11366, 11541, 11843, 11852,\\&&&
12285, 12729, 12735, 12784, 12899, 13010, 13163, 13295, 13522,\\&&&
13565, 13676, 13814, 13863, 13916, 13948, 14178, 14236, 14384,\\&&&
14524, 14612, 14658, 14715, 14840, 14914, 15003, 15195, 15259,\\&&&
15263, 15323, 15349, 15369, 15458, 15486, 15522, 15581, 15637,\\&&&
15708, 15891, 16053, 16104, 16114, 16538, 16633, 16720, 16735,\\&&&
16869, 17020, 17264, 17405, 17552, 17764, 17777, 17799, 17892,\\&&&
17950, 18167, 18393, 18529, 18529, 18629, 18663, 18856, 18953,\\&&&
19088, 19240, 19257, 19435, 19450, 19469, 19472, 19587, 19664,\\&&&
19770, 19950, 19963, 20358, 20434, 20518, 20581, 20735, 20754,\\&&&
20915, 21000, 21103, 21145, 21201, 21220, 21286, 21613, 21625,\\&&&
21668, 21924, 22010, 22075, 22313, 22323, 22429, 22482, 22598,\\&&&
22776, 23021, 23068, 23077, 23126, 23259, 23338, 23445, 23765,\\&&&
24014, 24262, 24317, 24511, 24795, 24887, 24917, 25059, 25091,\\&&&
25237, 25371, 25581, 25791, 25807, 25956, 26050, 26087, 26201,\\&&&
26361, 26405, 26433, 26700, 26791, 26840, 27090, 27115, 27204,\\&&&
27307, 27346, 27715, 27875, 28140, 28193, 28220, 28340, 28526,\\&&&
28786, 29033, 29062, 29245, 29293, 29637, 29752, 29757, 29918,\\&&&
30008, 30162, 30285, 30287, 30356, 30704, 30822, 30895, 30941,\\&&&
30945, 31017, 31123, 31230, 31292, 31341, 31355, 31423, 31557,\\&&&
31599, 31630, 32301, 32545, 32551, 32727\}\\
  \hline
\end{tabular}\\
\begin{tabular}{|c|c||c|l|}
  \hline
  $N$ & $d=|K|$ & $\epsilon$ & $K$ \\
  \hline
65536 & 257 & 0,0077 & \{121, 344, 358, 407, 1246, 1414, 1479, 1836, 2312, 2994,\\&&&
3003, 3400, 3767, 3887, 4428, 4633, 5018, 5976, 6199, 6515,\\&&&
6600, 6661, 7398, 7541, 7674, 7848, 8021, 8510, 9343, 9874,\\&&&
9940, 10210, 10831, 11136, 11218, 11483, 11727, 11737, 12151,\\&&&
12375, 12465, 12583, 12612, 13724, 13978, 14139, 14340, 14452,\\&&&
14545, 14670, 15223, 15311, 15409, 16335, 16410, 16696, 16804,\\&&&
16851, 16933, 16983, 17008, 17010, 17252, 17277, 17614, 17726,\\&&&
17831, 18134, 19016, 19489, 19665, 20222, 20707, 20823, 20856,\\&&&
21689, 21754, 21832, 21911, 21946, 22394, 22608, 22889, 23027,\\&&&
23669, 23775, 24299, 24524, 24589, 24622, 24661, 24763, 24792,\\&&&
25031, 25240, 25266, 25500, 25903, 26230, 26276, 26406, 26590,\\&&&
26639, 26696, 26791, 27007, 27317, 27380, 27470, 27791, 28094,\\&&&
28793, 28897, 28950, 29173, 29420, 29423, 29614, 30499, 30820,\\&&&
30993, 31400, 31635, 31670, 32239, 32424, 33261, 33606, 33836,\\&&&
33990, 34194, 34610, 34799, 35330, 35341, 35744, 35783, 35868,\\&&&
35873, 36187, 36293, 36381, 36569, 37143, 37156, 37172, 37435,\\&&&
38264, 38669, 38681, 38822, 39126, 39373, 39714, 39730, 39888,\\&&&
40764, 40951, 41008, 41488, 41979, 42749, 43311, 43814, 43866,\\&&&
44366, 44366, 44874, 45384, 46111, 46158, 46205, 46266, 46295,\\&&&
46389, 46476, 46551, 46806, 46819, 47136, 47395, 47644, 47722,\\&&&
48035, 48102, 48177, 48342, 48688, 49452, 49489, 49738, 49790,\\&&&
50066, 50499, 50876, 50922, 51086, 51230, 51588, 51886, 52012,\\&&&
52130, 52162, 52345, 52360, 52551, 53024, 53039, 53067, 53779,\\&&&
54154, 54211, 54265, 54267, 54642, 55419, 55786, 56173, 56953,\\&&&
57555, 57580, 57742, 58058, 58163, 58518, 58770, 58986, 59180,\\&&&
59709, 60051, 60126, 60421, 60958, 61502, 61671, 61725, 61759,\\&&&
61944, 61975, 62277, 62349, 62380, 62551, 63044, 63141, 63182,\\&&&
63206, 63468, 63718, 63982, 64328, 64465, 64883, 65054, 65189,\\&&&
65299, 65412\}\\
131072 & 257 & 0,0084 & \{772, 774, 2145, 2407, 2485, 2891, 3616, 4413, 4922, 5228,\\&&&
5394, 6708, 7237, 7252, 7298, 7313, 7373, 7953, 8324, 8637,\\&&&
9322, 9355, 10460, 11226, 11229, 13170, 13261, 14003, 14182,\\&&&
14304, 15260, 15454, 15835, 16167, 16693, 16826, 17635, 18344,\\&&&
18383, 18820, 19078, 19388, 19931, 20049, 20378, 20395, 20499,\\&&&
21851, 21993, 22904, 23095, 24082, 24160, 25794, 26133, 26180,\\&&&
27091, 27223, 27763, 29235, 29458, 30470, 30836, 30900, 31102,\\&&&
31288, 31608, 31676, 32182, 33107, 33655, 33669, 34081, 34411,\\&&&
35637, 36607, 38197, 38270, 38364, 39224, 39300, 39791, 39900,\\&&&
40200, 41026, 41723, 41785, 41785, 42556, 42731, 42998, 43298,\\&&&
44016, 44138, 44167, 44465, 44601, 46475, 47358, 48194, 48604,\\&&&
49222, 49309, 49471, 49496, 49917, 50859, 50933, 51429, 51914,\\&&&
53521, 54010, 54102, 54408, 56203, 56245, 56458, 56620, 57113,\\&&&
57636, 59239, 60584, 61504, 61718, 62340, 62390, 63065, 63604,\\&&&
64771, 66011, 66159, 66753, 67206, 68176, 68309, 69197, 69713,\\&&&
69812, 69968, 70778, 70794, 71422, 72485, 72784, 73141, 74134,\\&&&
74276, 74693, 75833, 76246, 76311, 76811, 77244, 77421, 78372,\\&&&
78768, 79700, 79978, 80285, 80632, 82124, 82195, 82320, 82381,\\&&&
83054, 83426, 83459, 83773, 83892, 84143, 84813, 85110, 85568,\\&&&
86191, 86764, 87463, 87521, 88214, 88695, 89287, 89735, 89913,\\&&&
91602, 92557, 92792, 93698, 94759, 95232, 95241, 96241, 96331,\\&&&
96435, 96857, 97153, 97574, 97579, 98013, 98017, 98550, 99068,\\&&&
99359, 99472, 99509, 99571, 99759, 100101, 100168, 100663,\\&&&
101576, 103598, 103731, 104217, 104460, 104505, 104681, 104863,\\&&&
105078, 105727, 105741, 105878, 106540, 106851, 107056, 107406,\\&&&
107461, 108799, 109336, 110194, 110273, 111420, 111658, 111691,\\&&&
111888, 113206, 114094, 114169, 114809, 115481, 117505, 118652,\\&&&
119680, 120271, 120542, 121922, 122455, 122725, 122972, 123240,\\&&&
124027, 127172, 127384, 129999, 130151, 130232, 130246, 130454,\\&&&
130667\}\\
262144 & 257 & 0,0087 & \{643, 786, 2493, 2792, 3952, 5002, 8047, 8732, 9863, 9923,\\&&&
9994, 10178, 11898, 17862, 18565, 18607, 21242, 21528, 22860,\\&&&
23091, 23124, 25805, 25936, 30386, 30450, 31930, 33157, 33728,\\&&&
34668, 34925, 36197, 37053, 40079, 40564, 41823, 42064, 42135,\\&&&
42204, 44349, 44421, 45726, 45819, 46250, 49446, 49735, 50089,\\&&&
50777, 51297, 51562, 51686, 53043, 53100, 53429, 55336, 56002,\\&&&
56042, 56243, 57381, 57450, 57995, 58535, 58715, 59177, 59475,\\&&&
59614, 59730, 61501, 62086, 64305, 66567, 67115, 67699, 70831,\\&&&
71530, 71955, 73421, 77485, 78660, 79272, 80828, 80873, 81590,\\&&&
83573, 84189, 84485, 84664, 87060, 88426, 88631, 91749, 93056,\\&&&
94112, 94281, 94364, 95113, 95198, 95735, 97564, 98515, 103543,\\&&&
103636, 104814, 107432, 109349, 109350, 111112, 111694, 112007,\\&&&
114685, 116651, 118571, 118812, 119696, 119965, 124385, 125981,\\&&&
126479, 126899, 127339, 128396, 129073, 129724, 129891, 130369,\\&&&
130911, 131593, 135007, 135297, 136521, 138159, 139128, 139478,\\&&&
139910, 140610, 141476, 142441, 142834, 145365, 145532, 146253,\\&&&
146567, 146572, 147010, 150108, 150727, 151550, 152876, 153080,\\&&&
156288, 156629, 158836, 160040, 160048, 160091, 160655, 161324,\\&&&
161811, 162257, 163422, 163607, 166151, 167293, 167994, 168688,\\&&&
169425, 170740, 171024, 171419, 172542, 172917, 175467, 175651,\\&&&
176271, 178336, 179124, 179446, 180684, 181874, 182932, 183345,\\&&&
187639, 188393, 188552, 188830, 189557, 189886, 190131, 192467,\\&&&
192833, 192861, 194443, 195085, 195862, 197149, 199574, 200955,\\&&&
202542, 203172, 204736, 208446, 209080, 209710, 210521, 210991,\\&&&
211848, 212169, 212405, 212946, 213156, 213397, 213555, 213831,\\&&&
214069, 214361, 217622, 218140, 218901, 219175, 219198, 219214,\\&&&
219486, 220757, 220812, 220819, 221229, 224600, 226841, 226966,\\&&&
227424, 228322, 229489, 229963, 232213, 232492, 233008, 233558,\\&&&
235125, 239310, 239612, 242526, 245405, 245595, 246231, 246503,\\&&&
249122, 252322, 252444, 253483, 256161, 257947, 259051, 259051,\\&&&
259537, 259615, 259951, 260758, 260983\}\\
  \hline
\end{tabular}\\
\begin{tabular}{|c|c||c|l|}
  \hline
  $N$ & $d=|K|$ & $\epsilon$ & $K$ \\
  \hline
524288 & 257 & 0,0094 & \{1615, 3276, 3850, 10222, 10748, 10831, 17750, 20687,\\&&&
22255, 25484, 28175, 33358, 34420, 35696, 36370, 41428, 41990,\\&&&
46012, 46846, 51097, 56166, 56696, 65322, 66528, 67490, 67532,\\&&&
67806, 73080, 76212, 77275, 77658, 78951, 81508, 81881, 82454,\\&&&
83581, 87248, 90272, 95060, 95991, 96331, 96838, 97920, 99331,\\&&&
106551, 107350, 110382, 112254, 112808, 112881, 119851, 122994,\\&&&
123235, 124878, 126992, 128131, 129997, 132026, 132228, 132653,\\&&&
133275, 138032, 139430, 139791, 141273, 143865, 152073, 152156,\\&&&
152303, 155348, 162508, 162527, 163778, 166648, 168688, 168935,\\&&&
169368, 181587, 182141, 183202, 185513, 186724, 188131, 189804,\\&&&
192210, 194193, 195027, 197874, 202220, 206548, 207206, 207767,\\&&&
208320, 208563, 209972, 211120, 220612, 220793, 221356, 222121,\\&&&
223202, 223371, 226785, 229157, 229314, 229559, 230930, 232951,\\&&&
235988, 247192, 248318, 253535, 255473, 257021, 258022, 260572,\\&&&
265789, 267305, 268742, 270793, 272224, 272723, 274159, 275810,\\&&&
276154, 279391, 279661, 279958, 281152, 281548, 281896, 282950,\\&&&
283519, 286717, 289336, 289384, 290557, 292341, 294333, 300365,\\&&&
301294, 301938, 304098, 304125, 304192, 304437, 304875, 305721,\\&&&
306050, 313199, 313643, 315660, 316031, 318610, 320155, 320791,\\&&&
323338, 325026, 326215, 336427, 338891, 339709, 340195, 340214,\\&&&
340847, 342789, 345044, 345564, 346582, 351118, 351629, 352393,\\&&&
357842, 358606, 360355, 360576, 363541, 364630, 367327, 367823,\\&&&
368259, 370047, 372143, 373555, 373652, 374375, 376993, 377869,\\&&&
379080, 383452, 384780, 386725, 394130, 394202, 394390, 396055,\\&&&
403615, 404352, 407655, 409360, 415326, 416204, 417553, 418104,\\&&&
418199, 419876, 423733, 425757, 429996, 431507, 431959, 434585,\\&&&
436795, 438137, 439145, 439207, 443686, 443764, 447885, 449709,\\&&&
449792, 456400, 461258, 463187, 466845, 467049, 467760, 467805,\\&&&
468085, 469795, 471935, 472031, 473809, 473974, 476249, 477114,\\&&&
477902, 478772, 483581, 485837, 489137, 489174, 490662, 490813,\\&&&
495086, 495795, 498906, 500436, 506276, 508742, 510350, 511087,\\&&&
511363, 512040, 516576, 523523, 524099\}\\
1048576 & 257 & 0,0099 & \{1467, 8352, 8816, 10832, 14043, 15376, 18119, 25830,\\&&&
27337, 32546, 37049, 37543, 45103, 54179, 67860, 68075, 68222,\\&&&
69594, 71134, 78516, 80251, 86576, 88394, 90601, 90755, 118896,\\&&&
120387, 124837, 128143, 128921, 130427, 133746, 138043, 138065,\\&&&
154516, 165415, 174585, 178251, 183858, 192691, 193847, 196056,\\&&&
196497, 197054, 211363, 213159, 213431, 216485, 216536, 227888,\\&&&
229597, 234076, 238281, 251990, 252438, 264721, 265129, 268479,\\&&&
269826, 275208, 282814, 283853, 286291, 287808, 292464, 306033,\\&&&
311587, 311977, 314512, 317470, 323161, 324128, 325290, 325908,\\&&&
340021, 340604, 342917, 350557, 360120, 365328, 366650, 368056,\\&&&
368681, 371016, 373137, 376943, 378240, 379580, 386183, 388267,\\&&&
389733, 397981, 401186, 401380, 403285, 403562, 409175, 412925,\\&&&
425315, 429225, 439091, 440128, 441974, 451491, 453350, 455257,\\&&&
463673, 465022, 470853, 471838, 473156, 473827, 477671, 483446,\\&&&
493524, 494683, 494909, 498627, 502340, 510007, 513598, 521346,\\&&&
521431, 523910, 523967, 526332, 528319, 529782, 537286, 538532,\\&&&
539991, 544570, 546715, 547703, 552112, 557102, 558204, 558325,\\&&&
562543, 565869, 568310, 577145, 578012, 581533, 589227, 589663,\\&&&
589775, 595664, 596046, 603022, 605813, 615224, 615924, 622248,\\&&&
630036, 636839, 639153, 640868, 644019, 645312, 648322, 649220,\\&&&
649408, 652215, 655253, 655921, 666619, 667786, 669245, 669849,\\&&&
670565, 676296, 677541, 684254, 687426, 688409, 690339, 692443,\\&&&
695302, 696777, 702652, 704205, 706731, 709802, 713815, 718962,\\&&&
721442, 730816, 734056, 740089, 747594, 755061, 761262, 761347,\\&&&
767787, 770110, 772491, 775670, 784503, 788361, 798914, 805442,\\&&&
807532, 810018, 813144, 816859, 823588, 825894, 831745, 832123,\\&&&
840311, 842701, 851452, 852164, 856838, 858765, 863913, 871321,\\&&&
874036, 890582, 894550, 901692, 903377, 904508, 906382, 911171,\\&&&
912719, 914082, 914202, 919244, 929393, 935928, 938474, 940352,\\&&&
951250, 955217, 955349, 958506, 965616, 966070, 972528, 974053,\\&&&
980049, 982697, 989151, 995352, 998881, 1004423, 1007093,\\&&&
1013239, 1014671, 1016722, 1028171, 1036547, 1045632, 1047199,\\&&&
1048198\}\\

  \hline
\end{tabular}
}

\section{Quantum Digital Signature Based on Quantum Hashing}

The proposed quantum hashing is a suitable one-way function for quantum digital
signature protocol from \cite{GC:2001:Quantum-Digital-Signatures} and below we
describe its basic structure modified for the specific hashing function.

To sign a single message bit $b$ Alice picks from $\{1,\ldots, L\}$ uniformly at random a pair of keys $(K_0, K_1)$.
This pair constitutes her \emph{private} key.

Using her private key pair Alice creates a sufficient numbers of public key
pairs
$$
(\ket{h_{K_0}}, \ket{h_{K_1}})
$$
and sends them to potential recipients. It can be easily verified that quantum
states in each pair are nearly orthogonal and thus distinguishable with high probability.

Now, given a message bit $b$, Alice sends it to Bob together with
a part of her private key $K_b$, which constitutes her
signature.

%

Finally, Bob, the recipient of a signed message, validates the signature by
``uncomputing'' $\ket{h_{K_b}}$ the same way it was created, i.e. by a sequence
of controlled rotations by the negative angles and the Hadamard transform. If
the signature is correct Bob will always obtain the all-zero state out of it.
Otherwise, the probability of error will be bounded by $\varepsilon$ due to
$\varepsilon$-collision resistance of the hashing function.

This protocol uses $O(\log\log{L})$ qubits for public keys
, where $L$ is a security level parameter and it should be chosen to deal with
the $1/L$ probability of guessing what the private key is by the possible
forger.

\paragraph{Acknowledgements.}

Research was supported by the Russian Fund for Basic Research (under the grants
11-07-00465, 12-01-31216).

\bibliography{references}

\end{document}